\documentclass[aps,prl,twocolumn,longbibliography,superscriptaddress]{revtex4-2}

\usepackage{libertine}
\usepackage{amsmath,amsfonts,amssymb,bm}
\usepackage{braket} 
\usepackage{float} 
\usepackage{bbold}

\usepackage{slashed}
\usepackage{enumerate}
\usepackage{enumitem} 
\usepackage[all]{xy} 

\usepackage{graphics}
\usepackage[dvipsnames]{xcolor}
\definecolor{myblue}{rgb}{0.0, 0.5, 1.0}
\definecolor{mygreen}{rgb}{0.0, 0.65, 0.31} 
\definecolor{myred}{rgb}{1.0, 0.13, 0.32} 
\usepackage[normalem]{ulem} 

\usepackage[colorlinks=true, urlcolor=violet, linkcolor=myblue, citecolor=red, hyperindex=true, linktocpage=true]{hyperref}
\usepackage[nameinlink,capitalise]{cleveref}

\usepackage{tikz}
\usepackage[perpage]{footmisc} 

\allowdisplaybreaks[4] 

\def\Im{\mathop{\mathsf{Im}}}
\def\Re{\mathop{\mathsf{Re}}}

\def\tr{\mathop{\mathrm{tr}}}
\def\Tr{\mathop{\mathrm{Tr}}}
\def\sgn{\mathop{\mathrm{sgn}}}

\begin{document}
\title{Composite Fermion Theory of Fractional Chern Insulator Stability}

\author{Xiaodong Hu}
\affiliation{Department of Material Science and Engineering, University of Washington, Seattle, WA 98195, USA}

\author{Ying Ran}
\affiliation{Department of Physics, Boston College, Chestnut Hill, Massachusetts 02467, USA}

\author{Di Xiao}
\affiliation{Department of Material Science and Engineering, University of Washington, Seattle, WA 98195, USA}
\affiliation{Department of Physics, University of Washington, Seattle, Washington 98195, USA}

\date{\today}

\begin{abstract}
	We develop a mean-field theory of the stability of fractional Chern insulators based on the dipole picture of composite fermions (CFs). We construct CFs by binding vortices to Bloch electrons and derive a CF single-particle Hamiltonian that describes a Hofstadter problem in the enlarged CF Hilbert space, with the trace-condition term emerging naturally in the small-$q$ limit as part of the CF Hamiltonian. Going beyond the small-$q$ limit, we apply our theory to twisted MoTe$_2$ and calculate its CF band structures. The resulting CF phase diagram matches closely with that from exact diagonalization, and the projected many-body wavefunctions achieve exceptionally high overlaps with the latter. Our theory provides both a microscopic understanding and a computationally efficient tool for identifying fractional Chern insulators.
\end{abstract}

\maketitle

\emph{Introduction.}---Fractional Chern insulators (FCIs) are lattice analogs of the fractional quantum Hall effect (FQHE), but realized in zero magnetic field. Proposed over a decade ago~\cite{tang2011high,neupert2011fractional,sheng2011fractional,sun2011nearly,regnault2011fractional,xiao2011interface}, FCIs have only recently been observed experimentally in twisted MoTe$_2$ bilayers~\cite{cai2023signatures,park2023observation,xu2023observation,kang2024evidence} and in pentalayer graphene on hBN~\cite{lu2024fractional}. Unlike conventional FQHE platforms, these moir\'e systems offer a wide range of tuning knobs---such as twist angle, gate voltage, and dielectric environment---enabling exploration of a much richer phase diagram~\cite{cai2023signatures,park2023observation,xu2023observation,kang2024evidence,lu2024fractional,waters2025chern,xu2025signatures}. Given this experimental progress, understanding the stability of FCIs has become an urgent issue.

To date, two main theoretical approaches have been pursued. Numerically, techniques such as exact diagonalization (ED) and density matrix renormalization group (DMRG) have been used to map out FCI phase diagrams \cite{repellin2020chern,li2021spontaneous,grushin2015characterization}, but these are computationally expensive and limited to small system sizes. Analytical approaches, in contrast, have primarily focused on the \emph{sufficient} conditions to realize FCI phases, mostly by mimicking the lowest Landau level (LLL) physics~\cite{estienne2023ideal}. Notable examples include the trace condition restoring the Girvin-MacDonald-Platzman (GMP) algebra~\cite{girvin1986magneto} in the small-$q$ limit~\cite{parameswaran2012fractional,parameswaran2013fractional,roy2014band,jackson2015geometric,ledwith2020fractional}, the ideal band condition yielding $\bm k$-holomorphic Bloch wavefunctions~\cite{mera2021kahler,wang2021exact,wang2023origin}, and the notion of vortexability extending the flux attachment picture in combination of band projection~\cite{ledwith2023vortexability,fujimoto2025higher}. While insightful, these criteria apply only to narrow regions of parameter space similar to the LLL. In fact, recent numerical studies have shown that deviating from such conditions can, in some cases, enhance the many-body gap~\cite{shi2025effects,emanuel2025unifying}. This motivates a central question: Can we develop a physically transparent and computationally efficient framework for diagnosing FCI stability beyond the ideal limit?

In this Letter, we present a quantitative theory of FCI stability based on the dipole picture of composite fermions (CFs)~\cite{read1994theory,pasquier1998dipole,read1998lowest,murthy2003hamiltonian,murthy2012hamiltonian,hu2024hyperdeterminants,shi2024quantum,lu2024fractional-CF}. We construct CFs on a lattice by binding Bloch electrons to vortices residing in their own LLL, and derive an effective CF Hamiltonian using the method of preferred-charge substitution~\cite{murthy2003hamiltonian,dong2020noncommutative,ma2022quantitative}. In the small-$q$ limit, the resulting Hamiltonian describes a Hofstadter problem in the enlarged CF Hilbert space, where both the scalar potential and the magnetic field are periodic (see also Ref.~\cite{shi2024quantum}).
The trace condition arises as a special case: when the electron band is flat and the trace condition is satisfied, the Hamiltonian reduces to the Aharonov-Casher (AC) form~\cite{aharonov1979ground,shi2024adiabatic,morales2024magic}, which guarantees the exact flatness of the lowest CF band. To quantitatively extract the CF band structure, we go beyond the small-$q$ limit and employ the GMP expansion~\cite{murthy2012hamiltonian,chamon2012magnetic}, in combination with our recently developed lattice projective construction of FCI states~\cite{hu2024hyperdeterminants}. The collapse of the FCI phase is then signaled by the closing of the CF band gap.

As a concrete example, we apply our theory to study FCI stability in twisted MoTe$_2$. Using only the CF single-particle Hamiltonian, we obtain a phase diagram that captures both the shape of the FCI region and the location of the maximal many-body gap, in agreement with ED results. We show that the dependence on Coulomb interaction strength and twist angle can be understood by examining the evolution of the CF band structure. We further validate our approach by constructing many-body electron wavefunctions through projection of the CF states, and find excellent overlaps with ED ground states. Our CF-based framework thus provides an efficient and analytically tractable tool for diagnosing FCI stability, offering predictive power and physical insight into the interplay between band geometry and interactions.

\emph{The Preferred CF Hamiltonian.}---Although not as widely known as the flux attachment formalism~\cite{jain1989composite,halperin1993theory}, the dipole picture offers a direct and physically motivated route to constructing the CF Hamiltonian. In the context of the FQHE, a CF dipole is formed by binding each electron to a $2s$-fold vortex~\cite{read1994theory}. Each vortex carries a charge opposite to that of the electron, $q_v = -c^2 q_e$, where $c^2 \equiv 2s \nu$ and $\nu$ is the electron filling factor. As a result, the total CF charge becomes $q_\text{CF} = (1 - c^2) q_e$. When $\nu$ lies in the Jain sequence $\nu = p / (2sp + 1)$, the CF filling factor becomes an integer $p$, and the FQHE of electrons maps to the integer quantum Hall effect of CFs. This construction underlies the extended Hamiltonian theory of Murthy and Shankar, which provides a powerful tool for studying the FQHE~\cite{murthy2003hamiltonian}. In what follows, we apply it to FCIs and present a self-contained derivation within this framework; readers may consult Ref.~\cite{murthy2003hamiltonian,hu2024hyperdeterminants} for more background.

Our starting point is the electronic Hamiltonian projected into a single band (we omit the band index for simplicity):
\begin{equation}\label{eq:single-band electronic Hamiltonian}
	H = H_0 + U = \sum_{\bm k} \varepsilon_{\bm k} c_{\bm k}^\dagger c_{\bm{k}} + \frac{1}{2} \sum_{\bm q} V_{\bm q} :\bar\rho_{e,\bm q} \bar\rho_{e,-\bm q}: \;,
\end{equation}
where $\varepsilon_{\bm{k}}$ is the band dispersion, $V_{\bm{q}}$ is the Coulomb interaction, and $\bar\rho_{e,\bm{q}} = \sum_{i=1}^{N} P_e e^{i \bm q \cdot \hat{\bm r}_{e,i}} P_e$ is the projected electron density operator, with $P_e = \sum_{\bm k} \ket{\psi_{\bm k}} \bra{\psi_{\bm k}}$ the Bloch band projector. We assume that the band carries Chern number $C = -1$ and the electron filling factor falls within the Jain sequence $\nu=p/(2sp+1)$.

To construct CFs on a lattice, we introduce vortices as auxiliary degrees of freedom similar to the FQHE. Although we continue to call them ``vortices'', they are best viewed as fictitious particles that carry fractional charge $q_v=-c^2 q_e$, and experience an effective magnetic field set by the average Berry curvature flux, $B_\text{eff} = h/(e A_\text{u.c.})$, where $A_\text{u.c.}$ is the unit cell area. The vortex unit cell, determined by magnetic translation symmetry, is enlarged by a factor of $(2sp+1)$ relative to the electron unit cell.

The CF is a bound state of an electron and a vortex that lives in the enlarged single-particle Hilbert space
\begin{equation}\label{eq:Hilbert space enlargement}
	\mathcal H_\text{CF} = \mathcal H_e \otimes \mathcal H_v \;,
\end{equation}
where $\mathcal H_e$ is spanned by the Bloch states and $\mathcal H_v$ by the vortex LLL states. The ``binding'' between a Bloch electron and a vortex is achieved by replacing the electron density operator with the so-called \emph{preferred charge} density operator~\cite{murthy2003hamiltonian,dong2020noncommutative,ma2022quantitative}, reflecting the dipole structure of CFs:
\begin{equation}\label{eq:preferred charge}
	\bar\rho_{e,\bm q} \mapsto \bar\rho^P_{\bm q} = \bar\rho_{e,\bm q} \otimes \mathbb{1}_v - c^2 \mathbb{1}_e \otimes \bar\rho_{v,-\bm q} \;,
\end{equation}
where $\bar\rho_{v,\bm q}$ is the LLL-projected vortex density operator:
\begin{equation}\label{eq:vorex density LLL projection}
	\bar\rho_{v,\bm q}=\sum_{i=1}^N P_v e^{-i\bm q\cdot\hat{\bm r}_{v,i}}P_v=e^{-(q\ell_v)^2/4}\sum_{i=1}^N e^{-i\bm q\cdot\hat{\bm R}_{v,i}} \;,
\end{equation}
with the vortex guiding center coordinates $\hat{\bm R}_v = P_v\hat{\bm r}_v P_v$, and the corresponding magnetic length $\ell_v = \sqrt{\hbar / |q_v B_\text{eff}|}$. The minus sign in the exponent of $\bar\rho_{v,\bm q}$ reflects the fact that the magnetic translation of a negatively charged vortex is equivalent to a reversed translation of the positive charge.

To carry out the preferred charge substitution, we first unpack the normal-ordered electron interaction in \cref{eq:single-band electronic Hamiltonian} and replace $\bar\rho_{e,\bm q}$ with $\bar\rho^P_{\bm q}$. We then re-normal order the resulting Hamiltonian in terms of $\bar\rho^P_{\bm q}$ to recover the standard form of an interacting Hamiltonian, arriving at (see Supplementary Material (SM)~\cite{Supplemental_Material} for details):
\begin{align}\label{eq:normal-ordered preferred Hamiltonian}
	H^P = H^P_\text{single} + H^P_\text{int}
	&= H_0 + H_0^P + H^P_\text{int} \;,
\end{align}
where $H_0$ is the electron band dispersion, $H^P_\text{int} = \sum_{\bm q} V_{\bm q} :\bar\rho^P_{\bm q} \bar\rho^P_{-\bm q}:$ is the normal-ordered CF interaction, and $H_0^P$ is a new single-particle term arising from the normal-ordering:
\begin{equation}\label{eq:second quantized H_0^P}
	H_0^P = \sum_{\bm k, \bm k'} \sum_{z_v, z_v'} H_0^P(\bm k, z_v; \bm k', z_v') \, c_{\bm k, z_v}^\dagger c_{\bm k', z_v'} \;.
\end{equation}
Here, $H_0^P$ is expressed in the CF basis composed of the electron Bloch states $\ket{\psi_{\bm k}}$ and the vortex coherent states defined by the complex guiding-center coordinate $\hat R_{v}|z_v\rangle=(\hat R_{v,x}+i\hat R_{v,y})|z_v\rangle=z_v|z_v\rangle$. Its matrix elements are
\begin{align}\label{eq:matrix element of the preferred Hamiltonian}
	H_0^P&(\bm k, z_v; \bm k', z_v') = \nonumber\\
	&\frac{-c^2}{2} \sum_{\bm q} V_{\bm q} \langle\psi_{\bm k},z_v|\bar\varrho_{e,\bm q} \otimes \bar\varrho_{v,\bm q} + \text{h.c.}|\psi_{\bm k'},z_v'\rangle \;,
\end{align}
where $\bar\varrho_{\alpha,\bm q} = e^{i\sgn(\alpha)\bm q\cdot\hat{\bm r}_\alpha}$ denotes the projected single-particle density operator of species $\alpha=e,v$ with $\sgn(e) = 1$ and $\sgn(v) = -1$.

\Cref{eq:normal-ordered preferred Hamiltonian,eq:second quantized H_0^P,eq:matrix element of the preferred Hamiltonian}
constitute the core result underlying our CF theory. They describe the band structure of CFs and provide the foundation for analyzing FCI stability. Crucially, at filling $\nu = p/(2sp+1)$, the CFs occupy exactly $p$ bands, mapping the problem of an FCI of electrons to that of an integer Chern insulator of CFs. This follows from the requirement that the CF periodicity must be compatible with both the electron and vortex unit cells, leading to an enlarged unit cell and hence $p$ filled CF bands.

In the remainder of this Letter, we demonstrate that much of the CF physics is already captured by the CF single-particle Hamiltonian $H^P_\text{single} = H_0 + H_0^P$, which encodes interaction effects through its dependence on the Coulomb potential $V_{\bm q}$.

\textit{Small-$q$ Limit and the Trace Condition}.---It is instructive to first consider the small-$q$ expansion of the projected density operators of both species $\alpha = e, v$:
\begin{equation}\label{eq:long wavelength limit expansion}
	\bar\varrho_{\alpha,\bm q} \simeq P_\alpha + i\, \sgn(\alpha)\, q_\mu P_\alpha \hat r_\alpha^\mu P_\alpha - \frac{1}{2} q_\mu q_\nu P_\alpha \hat r_\alpha^\mu \hat r_\alpha^\nu P_\alpha \;.
\end{equation}
Assuming further that $V_{\bm q}$ is isotropic, then only terms $\sim\bm q^2$ survive after integrating over $\bm q$. Up to a constant, the CF single-particle Hamiltonian takes the form (see SM~\cite{Supplemental_Material}):
\begin{equation}\label{eq:CF single particle Hamiltonian}
	\begin{split}
		H^P_{\text{single}}(\bm k,z_v) &=
		\varepsilon_{\bm k} + \frac{\hbar^2}{2m_*\ell_e^2\ell_v^2} (\tr \mathcal G_{\bm k} + \Omega_{\bm k}) \\
		&\quad + \frac{\hbar^2}{2m_*\ell_e^2\ell_v^2}(2\mathcal D - z_v)(2\bar{\mathcal D} - \hat{\bar z}_v) \;,
	\end{split}
\end{equation}
where $\ell_\alpha = \sqrt{\hbar/|q_\alpha B_\text{eff}|}$ and the CF effective mass is generated entirely by the Coulomb interaction:
\begin{equation}\label{eq:CF mass}
	\frac{\hbar^2}{2m_*\ell_e^2\ell_v^2} = \frac{c^2}{4} \sum_{\bm q} V_{\bm q} \bm{q}^2 e^{-\lambda^2 q^2} \;.
\end{equation}
We have added a $\lambda$-parameterized exponential regulator to account for the vortex LLL density operator form factor and the higher-order corrections from electron density operators (see SM~\cite{Supplemental_Material} for details).

The terms proportional to $1/m_*$ in Eq.~\eqref{eq:CF single particle Hamiltonian} reflect geometric properties of the underlying electron band. The complex covariant derivative is defined as $\mathcal D = \tfrac{1}{2}(\mathcal D_x - i\mathcal D_y)$, where $\mathcal D_\mu = i\partial_{k_\mu} + A_\mu(\bm k)$, and $A_\mu(\bm k) = i\braket{u_{\bm k}|\partial_{k_\mu}u_{\bm k}}$ is the Berry connection. The operator $\hat{\bar z}_v = -2\ell_v^2\partial_z$ is defined so that the vortex guiding-center coordinates satisfy $[\hat R_{v,x}, \hat R_{v,y}] = i \ell_v^2$. Interestingly, the trace condition expression $\tr \mathcal G_{\bm k} + \Omega_{\bm k}$, where $\mathcal G_{\bm k,\mu\nu} = \Re[\braket{\partial_{k_\mu} u_{\bm k} | \partial_{k_\nu} u_{\bm k}}] - A_\mu A_\nu$ is the quantum metric and $\Omega_{\bm k} = \Im[\braket{\partial_{k_x} u_{\bm k} | \partial_{k_y} u_{\bm k}}]$ is the Berry curvature, appears naturally in the CF Hamiltonian. Had we assumed the electron band carries positive Chern number $C = 1$, the sign of $\Omega_{\bm k}$ would be reversed.

We note that a Hamiltonian of the same form as \cref{eq:CF single particle Hamiltonian} was proposed in Ref.~\cite{shi2024quantum}, based on fitting a postulated quadratic electron-vortex dipole model. Our result provides a first-principle derivation of this structure as the small-$q$ limit within the preferred charge formalism.

\Cref{eq:CF single particle Hamiltonian} describes a Hofstadter problem in the $(\bm k, z_v)$ space: the first two terms act as a periodic scalar potential, while the last term, taking the form of an AC Hamiltonian~\cite{aharonov1979ground}, describes a periodic magnetic field. It is clear that if the electron band is flat ($\varepsilon_{\bm k} = 0$) and the trace condition $\tr \mathcal G_{\bm k} + \Omega_{\bm k} = 0$ is satisfied, the CF Hamiltonian reduces to a pure AC form, which guarantees the lowest CF band to be exactly flat. However, this trace condition alone does not determine the CF band gap, whose size controls the stability of the FCI phase as we show below. The CF gap arises from the combined effect of three factors: (i) the scalar-potential term (including band dispersion and the trace condition), (ii) the magnetic-field term associated with Berry-curvature, and (iii) the CF effective mass, which depends on interaction details.

\emph{CF Band Structure and CF Phase Diagram in tMoTe$_2$}---In the small-$q$ expansion~\cref{eq:CF single particle Hamiltonian}, we have introduced a regularization parameter for the effective mass. For a quantitative calculation of the CF band structure without any fitting parameter,  we will work with the CF Hamiltonian in~\cref{eq:matrix element of the preferred Hamiltonian} without taking the small-$q$ limit, where the information of the Berry curvature, quantum metric and CF effective mass are all implicitly encoded in the form factor.

To achieve this, we construct $N_s^2=(N_{e,1}\times N_{e,2})^2$ number of independent \emph{GMP operators} $\{\rho^{\text{GMP}}_{e,\bm q}\}$ from the matrix elements of electronic LLL density operator to satisfy the GMP algebra~\cite{girvin1986magneto}, and expand each $\bar\rho_{e,\bm q}$ following~\cite{murthy2012hamiltonian,hu2024hyperdeterminants}
\begin{equation}\label{eq:GMP expansion}
	\bar\rho_{e,\bm q} = \sum_{\bm G_{\text{std}}} c(\bm q, \bm G_{\text{std}}) \rho^\text{GMP}_{e,[\bm q] + \bm G_{\text{std}}} \;.
\end{equation}
Here $\sum_{\bm G_{\text{std}}}$ runs over $N_s$ number of reciprocal vectors, and $[\bm q]$ denotes $\bm q$ folded back to the first Brillouin zone (BZ). The coefficients (see SM \cite{Supplemental_Material} for derivation)
\begin{equation}
	c(\bm q, \bm G_{\text{std}}) = \frac{1}{N_s} \Tr\left[\bar\varrho_{e,\bm q} \varrho^\text{GMP}_{e,-[\bm q] - \bm G_{\text{std}}}\right]
\end{equation}
encodes \emph{all} lattice effects specific to the FCI. This so-called \emph{GMP expansion} is complete and can be used for any fermionic bilinear~\cite{murthy2012hamiltonian,chamon2012magnetic}. The resulting CF band structure problem can then be solved by diagonalizing the full-$q$ CF Hamiltonian.

\begin{figure*}[t!]
	\centering
	\includegraphics[width=0.9\textwidth]{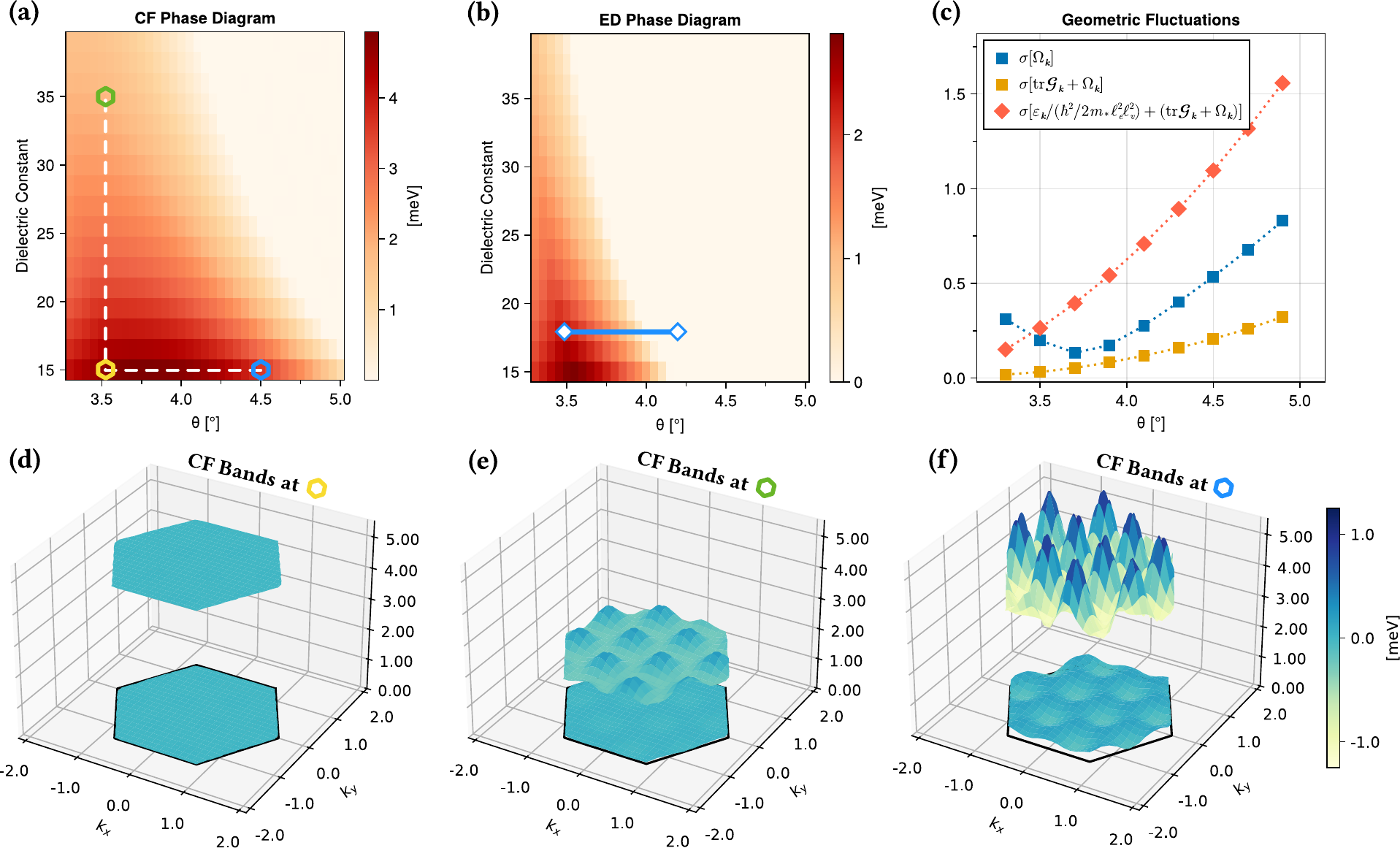}
	\caption{\textbf{CF phase diagram (a), $6 \times 4$ ED phase diagram (b), geometric fluctuations (c), and typical CF bands in twisted MoTe$_2$ (d--f).} The color maps in (a) and (b) represent the CF band gap and the many-body gap obtained from ED, respectively. Panel (c) shows the dimensionless standard deviation of the effective magnetic field $\Omega_{\bm k}$, traditional trace condition, and the effective scalar potential of the CF Hamiltonian~\eqref{eq:CF single particle Hamiltonian} at $\epsilon=15$, normalized by the moir\'e unit cell volume $A_{\text{u.c.}}$ and the effective CF mass $m_*\approx4.87$ meV measured from the calculated CF spectrum close to the ideal limit $\theta=3.55^\circ$. Panels (d)--(f) show the two lowest CF bands for three representative points labeled in (a), computed on a $9 \times 9$ electronic sample, with colors represents the band widths.
	Note: For visualization, the CF bands are unfolded into the electronic Brillouin zone (BZ), rather than plotted within the CF’s original BZ, which occupies only one-third of the electronic BZ at filling $\nu = 1/3$. The extra threefold degeneracy within CF's reduced BZ, known as the \emph{symmetry fractionalization pattern} due to the interplay between topological orders and magnetic translation symmetries~\cite{barkeshli2019symmetry,chen2017symmetry,essin2014spectroscopic,hu2024hyperdeterminants}, is also explicitly visible in our calculation.}
	\label{fig:combined_CF_phase_diagram_and_CF_bands}
\end{figure*}

We now apply our theory to the continuum model of the $R$-stacked twisted MoTe$_2$ bilayer. We assume the system is valley polarized, and focus on the effective Hamiltonian at the $K$ valley, given by~\cite{wu2019topological}:
\begin{equation}\label{eq:moire Hamiltonian}
	H_{K}=
	\begin{bmatrix}
		h_b(\bm r) & \Delta_T(\bm r) \\
		\Delta_T^\dagger(\bm r) & h_t(\bm r)
	\end{bmatrix},
\end{equation}
where $h_{b,t} = -(\bm k-\bm K_{b,t})^2/2m_\text{eff}+2v\sum_{i=1,3,5}\cos(\bm g_i\cdot\bm r\pm\psi)$ is the top/bottom layer Hamiltonian with a moir\'e potential, $\Delta_T = w(1+e^{-i\bm g_2\cdot\bm r}+e^{-i\bm g_3\cdot\bm r})$ is the interlayer tunneling, and $\bm g_i = \frac{4\pi}{\sqrt3 a/\theta}(\cos\frac{\pi(i-1)}{3},\sin\frac{\pi(i-1)}{3})$ are the moir\'e reciprocal vectors. The momenta $\bm K_b=\frac{1}{3}(\bm g_4+\bm g_5)$ and $\bm K_t=\frac{1}{3}(\bm g_3+\bm g_4)$ are the $K$ points of the top/bottom layer and $\theta$ is the twist angle. All parameters follows Ref.~\cite{wang2024fractional}: $m_*=0.6m_e$, $a=3.52$~\AA, $v=20.8~\text{meV}$, $\psi=107.7^\circ$, $w=-23.8~\text{meV}$. The Coulomb potential with dual-gate screening is taken to be $V_{\bm q}=\frac{1}{A_{\text{u.c.}}}\frac{e^2\tanh(|\bm q|d)}{2\epsilon\epsilon_0|\bm q|}$, where $d=300$~\AA is the gate-to-sample distance.

Let us focus on the hole filling $\nu=-2/3$ of the topmost valence band. We take a particle-hole transformation to convert it to an equivalent electron-doped model at filling $\nu=1/3$, squeezing out a Hartree-term $\varepsilon_{\bm k}^{\text{Hartree}}\equiv-\sum_{\bm G}V_{\bm G}(\sum_{\bm k'}\bar\varrho_{-\bm G}([\bm k']))\bar\varrho_{\bm G}([\bm k])$ that is absorbed into the original band dispersion $\varepsilon_{\bm k}\mapsto -\varepsilon_{\bm k}-\varepsilon_{\bm k}^{\text{Hartree}}$. We construct the CF Hamiltonian using the preferred charge substitution~\eqref{eq:preferred charge}, and solve it via the GMP expansion~\eqref{eq:GMP expansion} and our recently developed lattice projective construction~\cite{hu2024hyperdeterminants} \emph{without} resorting to the small-$q$ limit.

\Cref{fig:combined_CF_phase_diagram_and_CF_bands}(d) shows the CF band structure at $\theta = 3.55^\circ$ and $\epsilon = 15$. Only the two lowest CF bands are shown, since at $\nu = 1/3$ only the lowest band is fully filled, with all higher bands empty. We observe that both bands are flat, reflecting the near-ideal geometric conditions of the underlying Bloch band: the trace condition is nearly saturated, and the fluctuations in Berry curvature is small [see \cref{fig:combined_CF_phase_diagram_and_CF_bands}(c)]. The electron band dispersion is also small. In this case, the CF Hamiltonian \cref{eq:CF single particle Hamiltonian} describes motion of a charged particle in a uniform magnetic field in the enlarged ($\bm k, z_v)$ space, and the CF bands closely resemble the flat CF Landau levels in the FQHE.

In the non-interacting CF limit, the CF band gap provides a direct estimate of the activation energy---the cost of creating a well-separated quasiparticle–quasihole pair. We plot this gap as a function of twist angle $\theta$ and dielectric constant $\epsilon$ in \cref{fig:combined_CF_phase_diagram_and_CF_bands}(a). The resulting CF phase diagram shows close agreement with ED results in \cref{fig:combined_CF_phase_diagram_and_CF_bands}(b), capturing not only the shape of the FCI region but also the location of the maximal many-body gap~\footnote{Note that when comparing the CF phase diagram with the ED phase diagram, one should keep in mind that the CF gap corresponds to a charge excitation but the ED gap measures the magnetoroton minimum~\cite{girvin1986magneto}.}. This agreement strongly supports the predictive power of our CF-based approach and confirms that the CF band gap serves as a reliable proxy for FCI stability.

To understand the breakdown of the FCI phase, we track the evolution of the CF band structure along the vertical line in \cref{fig:combined_CF_phase_diagram_and_CF_bands}(a), varying $\epsilon$ at fixed $\theta=3.55^\circ$. In FQHE, neglecting Landau level mixing, the many-body gap scales with the interaction strength and remains open until the interaction is completely turned off. FCIs, however, behaves differently. As $\epsilon$ increases, the inverse CF mass $1/m_*$ decreases [see~\cref{eq:CF mass}], leading to two key effects illustrated in \cref{fig:combined_CF_phase_diagram_and_CF_bands}(e). First, the second CF band shifts downward, reflecting the expected reduction in interaction energy. Second, the weakened interaction amplifies the influence of the electron band dispersion $\varepsilon_{\bm{k}}$, causing the CF bands to become increasingly dispersive. Once $\epsilon$ exceeds a critical value, the top and bottom CF bands touch and the gap closes, signaling the breakdown of the FCI phase. Notably, this transition cannot be captured by geometric criteria such as ideal-band condition or vortexability, which remain unchanged along this trajectory. In contrast, the electron band dispersion enters naturally into the CF Hamiltonian, allowing our framework to account for this interaction-driven destabilization.

Next, we fix the dielectric constant and vary the twist angle $\theta$. As shown in \cref{fig:combined_CF_phase_diagram_and_CF_bands}(c), both the scalar potential [the first two terms in \cref{eq:CF single particle Hamiltonian}] and the effective magnetic field $\Omega_{\bm k}$ exhibit increasing fluctuations with larger $\theta$. As a result, CF bands develop significant dispersion, as shown in \cref{fig:combined_CF_phase_diagram_and_CF_bands}(f) for $\theta = 4.5^\circ$. In contrast, the band center shifts only slightly, reflecting the fact that the effective mass remains approximately constant. Thus, the enhanced fluctuations in the CF band energy are the primary cause of the decreasing CF band gap at larger twist angles.

\begin{figure}[htb!]
	\centering
	\includegraphics[width=0.95\linewidth]{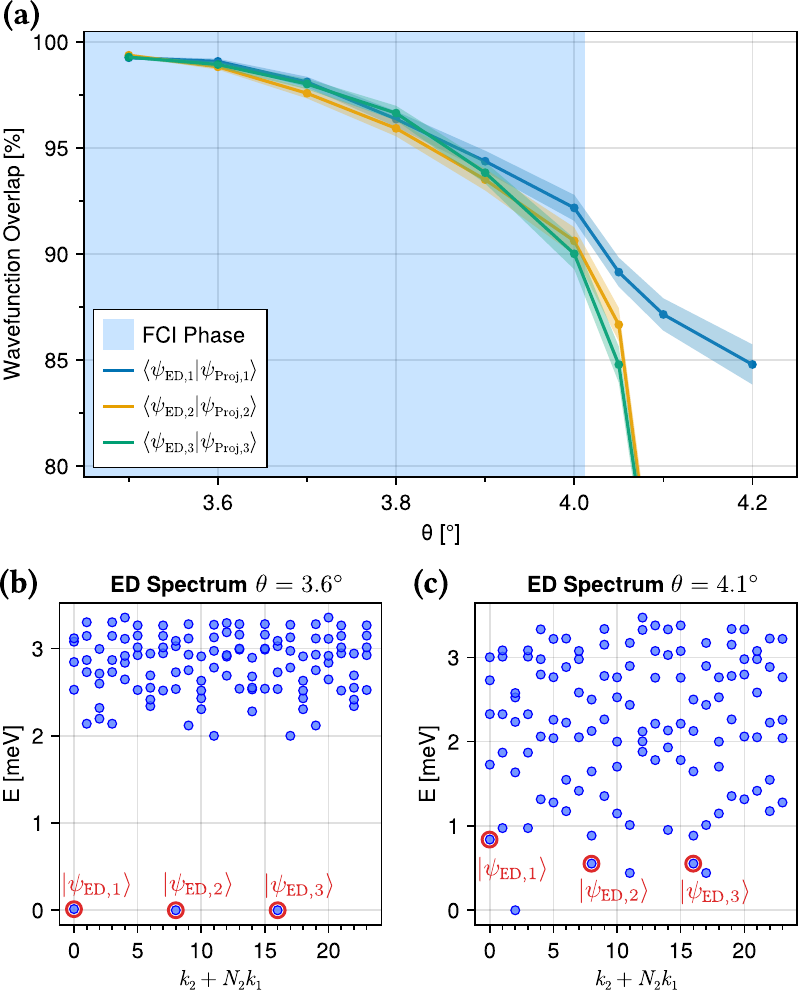}
	\caption{\textbf{Wavefunction overlap with ED results (a) and typical $6 \times 4$ ED spectra (b–c) before and after FCI breakdown in twisted MoTe$_2$}. Data are shown along the blue line cut in Fig.\ref{fig:combined_CF_phase_diagram_and_CF_bands}(b), spanning twist angles $\theta = 3.5^\circ$ to $4.2^\circ$ at dielectric constant $\epsilon = 18$, crossing the FCI phase boundary near $\theta_c \approx 4.01^\circ$.
	The projected wavefunction is constructed via \cref{eq:wavefunction overlap}, using CF states $\ket{\psi_\text{CF}}$ obtained from the single-particle Hamiltonian $H^P_\text{single}$. We compute the overlap between the projected wavefunction and the three ED eigenstates highlighted by red circles in (b) and (c), located at the center-of-mass momenta expected for the $6 \times 4$ FCI ground states. Shaded bands in (a) indicate statistical errors from variational Monte Carlo samplings.}
	\label{fig:proj_overlap}
\end{figure}

\emph{Physical Wavefunction from Projection}.---Our CF theory is formulated in an enlarged Hilbert space that includes auxiliary vortex degrees of freedom.
To obtain the physical electronic wavefunction, we must project out these unphysical degrees of freedom. This projection is implemented as~\cite{shi2024quantum,hu2024hyperdeterminants}:
\begin{equation}\label{eq:wavefunction overlap}
	\ket{\psi_\text{phys}} = \braket{\psi_v|\psi_\text{CF}},
\end{equation}
where $\ket{\psi_\text{CF}}$ is a Slater determinant of CF orbitals, $\ket{\psi_v}$ is the bosonic vortex state, and $\ket{\psi_\text{phys}}$ is the resulting physical state of electrons. While the choice of $\ket{\psi_v}$ is formally a gauge choice within the projective construction, a natural choice is the bosonic Laughlin state at filling $1/2s$: $\ket{\psi_v} = \ket{\psi_{\nu=1/2s,v}^\text{Laughlin}}$. This choice reproduces Jain's wavefunctions in the LLL limit and has been validated in both continuum and lattice settings~\cite{hu2024hyperdeterminants,shi2024quantum}.

To further test our framework, we numerically solve the CF single-particle Hamiltonian $H^P_\text{single}$, form a Slater determinant from the resulting orbitals, and project out the vortex degrees of freedom via \cref{eq:wavefunction overlap}. The procedure for carrying out this projection on a lattice was developed in Ref.~\cite{hu2024hyperdeterminants}, and involves evaluating a \emph{hyperdeterminant} \cite{gelfand1994hyperdeterminants,barvinok1995new,hillar2013most} of certain fusion tensors. The overlap of the resulting state $\ket{\psi_\text{phys}}$ and the exact ED ground state is then computed using Monte Carlo integration.

As shown in \cref{fig:proj_overlap}(b), at filling $\nu = 1/3$, there are three nearly degenerate ground states characterized by evenly spaced center-of-mass momenta~\cite{oshikawa2000commensurability,lu2020filling}. We generate the projected electronic states at these momenta and plot their overlap with the ED ground states in \cref{fig:proj_overlap}(a). The agreement is remarkable: within the FCI phase, the projected states achieve overlaps exceeding 90\%, with a peak value of $(99.37 \pm 0.06)\%$ at $\theta = 3.5^\circ$, where the many-body gap reach its maximum. As $\theta$ increases, the eigenstates outside the topological manifold descend in energy and eventually become the new ground states, signaling the breakdown of the FCI phase [\cref{fig:proj_overlap}(c)]. However, we continue to track the ED eigenstates at the three original momenta and find that their overlaps with the projected CF states remain substantial even after they are no longer the ground states. In contrast, the new ground states have negligible overlap with the three projected CF states. This result confirms that our framework not only qualitatively reproduces the FCI phase diagram, but also provides quantitatively accurate many-body wavefunctions through the electronic projection defined in~\cref{eq:wavefunction overlap}.

\emph{Summary}.---We have developed a mean-field theory of FCI phase stability based on the dipole picture of CFs. By introducing vortices as auxiliary degrees of freedom and implementing the preferred charge substitution, we derived a CF single-particle Hamiltonian that describes a Hofstadter problem in the enlarged CF Hilbert space. In the small-$q$ limit, the trace condition emerges naturally from this construction as the criterion for flatness of the lowest CF band. The calculated CF phase diagram closely matches that from ED, and the projected many-body wavefunctions achieve high overlap with ED results. This agreement strongly supports the validity of our microscopic framework. Moreover, the CF band structure is obtained by diagonalizing a single-particle Hamiltonian, making our approach both reliable and efficient for predicting FCI phases.

In this work, we have neglected the residual CF interactions $H^P_\text{int}$, in part because the preferred charge substitution already incorporates essential interaction effects---most notably, the CF effective mass is entirely Coulomb-driven. An important next step is to include $H^P_\text{int}$ to study collective excitations such as the magnetoroton mode. Another promising direction is to explore whether residual CF interactions can drive symmetry-breaking phases, such as CF charge orders or CF pairings, beyond the simple band picture. Additionally, generalizing our framework to composite fermi liquid (CFL) states in FCI~\cite{dong2023composite,goldman2023zero}, and incorporating band-mixing effects~\cite{yu2024fractional} remain to be important tasks. These extensions could further deepen our understanding of FCI physics and strengthen the connection between microscopic theory and experimental observations.

\emph{Acknowledgment}.---We thank Junkai Dong, Zhihuan Dong, Manato Fujimoto, Junren Shi, Jingtian Shi, and Jie Wang for stimulating discussions, and Junren Shi for sharing unpublished result. This work is supported by the Center on Programmable Quantum Materials, an Energy Frontier Research Center funded by DOE BES under Award No. DE-SC0019443. This research used resources of the National Energy Research Scientific Computing Center, a DOE Office of Science User Facility supported by the Office of Science of the U.S. Department of Energy under Contract No. DE-AC02-05CH11231 using NERSC award BES-ERCAP0033507.

\bibliography{ref}

\end{document}


\title{Supplemental Materials For ``Composite Fermion Theory of Fractional Chern Insulator Stability''}

\author{Xiaodong Hu}
\affiliation{Department of Material Science and Engineering, University of Washington, Seattle, WA 98195, USA}

\author{Ying Ran}
\affiliation{Department of Physics, Boston College, Chestnut Hill, Massachusetts 02467, USA}

\author{Di Xiao}
\affiliation{Department of Material Science and Engineering, University of Washington, Seattle, WA 98195, USA}
\affiliation{Department of Physics, University of Washington, Seattle, Washington 98195, USA}

\maketitle

\tableofcontents

\section{CF Preferred Hamiltonian for FQHE}
In this section we derive the single-particle Hamiltonian of composite fermions (CFs) for the fractional quantum Hall effect (FQHE) using the preferred charge substitution~\cite{murthy2003hamiltonian,dong2020noncommutative,ma2022quantitative}. Our starting point is the many-body Hamiltonian projected into the lowest Landau level (LLL):
\begin{equation}\label{eq:LLL Hamiltonian}
	H = \frac{1}{2}\sum_{\bm q} V_{\bm q} :\bar\rho_{e,\bm q} \bar\rho_{e,-\bm q}: \;,
\end{equation}
where $V_{\bm q}$ is the Coulomb interaction and
\begin{equation}
	\bar\rho_{e,\bm q}\equiv\sum_{i=1}^N\P_{e,\text{LLL}} e^{i\bm q\cdot\hat{\bm r}_{e,i}}\P_{e,\text{LLL}}=e^{-(\bm q\ell_e)^2/4} \sum_{i=1}^N e^{i\bm q\cdot\hat{\bm R}_{e,i}}
\end{equation}
is the electron density operator projected into the LLL. Note: throughout this SM and the main text, we will always use $\rho_{\bm q}$ to denote the second quantized density operators, or the many-body operator in the first quantized fashion, and use its variant $\varrho_{\bm q}\equiv e^{-i\bm q\cdot\hat{\bm r}}$ to denote the single-particle density operator, i.e., $\rho_{\bm q}\equiv\sum_{i=1}^N \varrho_{\bm q}^{(i)}$.

Here $\bm R_e$ is the electron guiding center, satisfying the commmutation relation
\begin{equation}
	[R_{e,x}, R_{e,y}] = -i\ell_e^2 \;,
\end{equation}
where $\ell_e = \sqrt{\hbar/eB}$ is the electron magnetic length.  Note that $\bar\rho_{e,\bm q}$ includes the LLL form factor.

To construct the CFs, we attach a $2s$-fold vortex to each electron.  Each vortex carries charge $q_v=-2s\nu e = -c^2e$, where $\nu$ is the electron filling factor, and experiences the same magnetic field as the electrons.  The vortex guiding center $\bm R_v$ satisfies the commutation relation
\begin{equation}
	[R_{v,x}, R_{v,y}] = i\ell_v^2 \;,
\end{equation}
where $\ell_v = \sqrt{\hbar/c^2eB} = \ell_e/c$ is the vortex magnetic length.

The binding of the electron and vortex is achieved by the so-called preferred charge substitution, in which we replace the projected electron density operator $\bar\rho_{e,\bm q}$ with the preferred charge density operator $\bar\rho_{\bm q}^P$:
\begin{equation}\label{eq:preferred charge substitution (general)}
	\bar\rho_{e,\bm q}\mapsto \bar\rho_{\bm q}^P\equiv\bar\rho_{e,\bm q}-c^2\bar\rho_{v,-\bm q},
\end{equation}
where the vortex density operator is projected into its own LLL:
\begin{equation}
	\bar\rho_{v,\bm q}\equiv\sum_{i=1}^N\P_{v,\text{LLL}} e^{-i\bm q\cdot\hat{\bm r}_{v,i}}\P_{v,\text{LLL}}\equiv e^{-(\bm q\ell_v)^2/4} \sum_{i=1}^N e^{-i\bm q\cdot\hat{\bm R}_{v,i}} \;.
\end{equation}
As emphasized by Shankar in Ref.~\cite{shankar1999hamiltonian}, the filling-dependent coefficient $-c^2$ in~\cref{eq:preferred charge substitution (general)} is crucial for satisfying Kohn's theorem in the LLL, ensuring the \emph{quartic} behavior of the static structure factor $S(\bm q)\sim(\bm q\ell_e)^4$ in the long wavelength limit.

To implement the substitution $\bar\rho_{e,\bm q} \to \bar\rho^P_{\bm q}$, we first unpack the normal-ordering in~\cref{eq:LLL Hamiltonian}. Generally, for fermionic bilinear operators $\bm A$ and $\bm B$, unpacking normal-ordering yields an additional single-particle term:
\begin{equation}\label{eq:removing normal-ordering}
	\bm A\bm B-:\bm A\bm B:\equiv\sum_n\hat A_n\hat B_n,
\end{equation}
which is a constant for the LLL-projected electron density operators. The resulting Hamiltonian after the substitution is the preferred Hamiltonian acting on the enlarged CF Fock space:
\begin{equation}
	H = \frac{1}{2}\sum_{\bm q} V_{\bm q}\bar\rho_{e,\bm q} \bar\rho_{e,-\bm q}\mapsto H^P = \frac{1}{2}\sum_{\bm q} V_{\bm q} \bar\rho^P_{\bm q} \bar\rho^P_{-\bm q}.
\end{equation}
To recover the standard form of the density-density Coulomb interactions between CFs, we need to re-normal order this preferred Hamiltonian, arriving at
\begin{equation}
	H^P = H^P_{\text{single}} + H^P_{\text{int}},\qquad H^P_{\text{int}} = \frac{1}{2}\sum_{\bm q} V_{\bm q}:\bar\rho^P_{\bm q} \bar\rho^P_{-\bm q}:,
\end{equation}
with an extra single-particle term $H_\text{single}^P$ squeezed out following~\cref{eq:removing normal-ordering}. In the first-quantized language,
\begin{align}
	H^P_{\text{single}} &= \frac{1}{2}\sum_{\bm q}V_{\bm q} \sum_n \bigg\{\bar\varrho_{e,\bm q}^{(n)}-c^2\bar\varrho_{v,-\bm q}^{(n)}\bigg\}\bigg\{\bar\varrho_{e,-\bm q}^{(n)}-c^2\bar\varrho_{v,\bm q}^{(n)}\bigg\}\nonumber\\
	&= \frac{1}{2}\sum_{\bm q}V_{\bm q}\sum_n \bigg\{e^{-(\bm q\ell_e)^2/4} e^{i\bm q\cdot\hat{\bm R}_{e,n}}-c^2 e^{-(\bm q\ell_v)^2/4} e^{i\bm q\cdot\hat{\bm R}_{v,n}}\bigg\}\bigg\{e^{-(\bm q\ell_e)^2/4}e^{-i\bm q\cdot\hat{\bm R}_{e,n}}-c^2 e^{-(\bm q\ell_v)^2/4}e^{-i\bm q\cdot\hat{\bm R}_{v,n}}\bigg\}.
\end{align}
Clearly only the e-v mixed term contributes, leading to
\begin{equation}
	H^P_{\text{single}} = \frac{-c^2}{2}\sum_{\bm q}V_{\bm q}e^{-\bm q^2(\ell_e^2+\ell_v^2)/4}\sum_n 2\cos\bigg[\bm q\cdot(\hat{\bm R}_{e,n}-\hat{\bm R}_{v,n})\bigg] \;.
\end{equation}

Next we follow Murthy and Shankar~\cite{murthy2003hamiltonian} to introduce the CF guiding center $\bm R$ and cyclotron coordinate $\bm\eta$:
\begin{equation}\label{eq:CF substution}
	\hat{\bm R} = \frac{\hat{\bm R}_e - c^2\hat{\bm R_v}}{1-c^2} \;, \qquad
	\hat{\bm \eta} = \frac{c}{1-c^2}(\hat{\bm R}_v - \hat{\bm R}_e) \;.
\end{equation}
They satisfy the commutation relations
\begin{equation}
	[R_x, R_y] = -i\ell_\text{CF}^2 \;, \qquad
	[\eta_x, \eta_y] = i\ell_\text{CF}^2 \;,
\end{equation}
where $\ell_\text{CF} = \ell_e/\sqrt{1-c^2}$ is the CF magnetic length.  In terms of the CF operators, the Hamiltonian reads
\begin{equation}
	H^P_\text{single} = \frac{-c^2}{2}\sum_{\bm q}V_{\bm q}e^{-\bm q^2(\ell_e^2+\ell_v^2)/4}\sum_n 2\cos\bigg[\bm q\cdot\left(c-\frac{1}{c}\right)\hat{\bm\eta}_n\bigg] \;.
\end{equation}

We can expand the cosine function to second order in $\bm q$.  For isotropic interaction $V_{\bm q}$, the angular integral over $\bm q$ gives a simple quadratic Hamiltonian:
\begin{equation}\label{eq:CF single-particle Hamiltonian (LLL)}
	H^P_\text{single} = \frac{\hbar^2}{2m_*\ell_\text{CF}^4}\sum_n\hat{\bm\eta}^2_n \;,
\end{equation}
where
\begin{equation}
	\frac{\hbar^2}{2m_*} = \frac{\ell_e^4}{4}\sum_{\bm q} V_{\bm q} e^{-(\bm q\ell_e)^2(1+1/c^2)/4} \bm q^2\;.
\end{equation}
It is clear that this Hamiltonian describes a CF cyclotron motion with frequency $\omega_c=q_{\text{CF}}B/m_*$, and a CF Landau level structure with spacing $\hbar\omega_c$.
If we insist on writing $H^P_\text{single}$ in terms of $\hat{\bm R}_e$ and $\hat{\bm R}_v$, then we have
\begin{equation}\label{eq:CF effective mass (LLL)}
	H_\text{single}^P = \frac{\hbar^2}{2m_*\ell_e^2\ell_v^2}\sum_n(\bm R_{e,n} - \bm R_{v,n})^2
\end{equation}

We see that the effective mass of the CF is entirely due to interaction.  The exponential term in the integral~\cref{eq:CF effective mass (LLL)} contains a regulator originating from the LLL form factors from both species in the long wavelength limit. But for general case (such as FCI, see below) this can be represented by a generic regulator $e^{-\lambda^2\bm q^2}$ as in the main text.

\section{CF Preferred Hamiltonian for FCI}
The derivation of the CF preferred Hamiltonian for fractional Chern insulators (FCIs) closely parallels that for the FQHE. We start from the single-band projected electronic Hamiltonian
\begin{equation}
	H\equiv H_0+U\equiv\sum_{\bm k}\varepsilon_{\bm k}c_{\bm k}^\dagger c_{\bm k} + \frac{1}{2}\sum_{\bm q}V_{\bm q}:\bar\rho_{e,\bm q}\bar\rho_{e,-\bm q}:.
\end{equation}
where $\bar\rho_{e,\bm{q}} = \sum_{i=1}^{N} \P_e e^{i \bm q \cdot \hat{\bm r}_{e,i}} \P_e$ is the projected electron density operator, with $\P_e = \sum_{\bm k} \ket{\psi_{\bm k}} \bra{\psi_{\bm k}}$ the Bloch band projector.

Unpacking the normal-ordering following~\cref{eq:removing normal-ordering} gives
\begin{align}
	H &= \sum_{\bm k}\bigg(\varepsilon_{\bm k}-\frac{1}{2}\sum_{\bm q}V_{\bm q}\varrho_{\bm q}([\bm k-\bm q])\varrho_{-\bm q}([\bm k])\bigg) c_{\bm k}^\dagger c_{\bm k} + \frac{1}{2}\sum_{\bm q}V_{\bm q}\bar\rho_{e,\bm q}\bar\rho_{e,-\bm q},\nonumber\\
	&\equiv\sum_{\bm k}(\varepsilon_{\bm k}-\varepsilon^V_{\bm k}) c_{\bm k}^\dagger c_{\bm k} + \frac{1}{2}\sum_{\bm q}V_{\bm q}\bar\rho_{e,\bm q}\bar\rho_{e,-\bm q},
\end{align}
where $\bar\varrho_{e,\bm q}([\bm k]) \equiv \langle\psi_{[\bm k+\bm q]}|\hat\rho_{e,\bm q}|\psi_{\bm k}\rangle$ is the matrix element of the single-band projected electron density.

Next, we enlarge the electronic Hilbert space by introducing the vortex degrees of freedom,
and apply the preferred-charge substitution~\cref{eq:preferred charge substitution (general)}.
The vortices are assumed to reside in their own LLL.
After re-normal-ordering the resulting Hamiltonian in terms of $\rho_{\bm q}^P$, we obtain the preferred Hamiltonian
\begin{equation}
	H^P \equiv H^P_{\text{single}} + H^P_{\text{int}},\quad H^P_{\text{int}}\equiv\frac{1}{2}\sum_{\bm q}V_{\bm q}:\rho^P_{\bm q}\rho^P_{-\bm q}:
\end{equation}
with the CF single-particle Hamiltonian given by
\begin{equation}\label{eq:CF single-particle Hamiltonian (without simplification)}
	H^P_{\text{single}}\equiv\sum_{\bm k}(\varepsilon_{\bm k}-\varepsilon^V_{\bm k}) c_{\bm k}^\dagger c_{\bm k} + \frac{1}{2}\sum_{\bm k,z_v}\sum_{\bm k',z_v'}\sum_{\bm q}V_{\bm q} \Big\langle\psi_{\bm k'},z_v'\Big|\Big(\bar\varrho_{e,\bm q}\otimes\mathbb1 - c^2\mathbb1\otimes\bar\varrho_{v,-\bm q}\Big)\Big(\text{h.c.}\Big)\Big|\psi_{\bm k},z_v\Big\rangle c_{\bm k',z_v'}^\dagger c_{\bm k,z_v}.
\end{equation}
Here the CF basis $\{|\psi_{\bm k},z_v\rangle\}$ is chosen to be composed of electronic Bloch state $|\psi_{\bm k}\rangle$ and vortex coherent state defined by the vortex guiding-center coordinates $\hat R_v|z_v\rangle\equiv(\hat R_{v,x}+i\hat R_{v,y})|z_v\rangle:=z_v|z_v\rangle$. Note that the eigenvalue of our coherent state differs by a factor of $\sqrt{2}\ell_v$ to the standard definition.

Clearly the previous shift to the dispersion $-\varepsilon^V_{\bm k}$ is exactly canceled by the e-e term in the second part of~\cref{eq:CF single-particle Hamiltonian (without simplification)}, and the v-v term is simply a constant. The only non-trivial contributions are from the e-v and v-e mutual terms:
\begin{equation}\label{eq:CF single-particle Hamiltonian (full)}
	H^P_{\text{single}} \equiv H_0 + H^P_0 \equiv \sum_{\bm k}\varepsilon_{\bm k} c_{\bm k}^\dagger c_{\bm k} + \frac{-c^2}{2}\sum_{\bm k,z_v}\sum_{\bm k',z_v'}\sum_{\bm q}V_{\bm q}\Big\langle\psi_{\bm k'},z_v'\Big|\bar\varrho_{e,\bm q}\otimes\bar\varrho_{v,\bm q}+\bar\varrho_{e,-\bm q}\otimes\bar\varrho_{v,-\bm q}\Big|\psi_{\bm k},z_v\Big\rangle c_{\bm k',z_v'}^\dagger c_{\bm k,z_v},
\end{equation}

Next let us work out the matrix element $H^P_0(\bm k,z_v;\bm k',z'_v)$. For analytical insight, we expand both density operators in the long-wavelength limit. The vortex density operator, due to the LLL projection by construction, already includes a form factor $e^{-(\bm q\ell_v)^2/4}$, which allows a small-$q$ expansion:
\begin{equation}\label{eq:LLL-projected vortex density operator}
	\bar\varrho_{v,\bm q} \simeq e^{-(\bm q\ell_v)^2/4}\P_{v,\text{LLL}}\bigg[1 - iq_\mu \hat R_{v,\mu} + \frac{-1}{2} \hat R_v^\mu R_v^\nu\bigg] + \mathcal O((\bm q\ell_v)^3).
\end{equation}
The electron operator, in contrast, generally lacks such exponentially decaying form factor under the band projection. However, we can still pull out a Gaussian term identical to that arising from the LLL projection of the electron density operator, at least \emph{in the small-$q$ limit}, so that
\begin{align}
	\bar\varrho_{e,\bm q}\equiv \P_e \varrho_{e,\bm q}\P_e &\simeq \P_e + iq_\mu \P_e \hat r_e^\mu \P_e + \frac{-1}{2} q_\mu q_\nu \P_e \hat r_e^\mu \hat r_e^\nu \P_e + \mathcal O((\bm q\ell_e)^3) \nonumber\\
	&\simeq e^{-(\bm q\ell_e)^2/4}\bigg[\P_e\left(1+\frac{1}{4}(\bm q\ell_e)^2\right) + iq_\mu \P_e \hat r_e^\mu \P_e + \frac{-1}{2} q_\mu q_\nu \P_e \hat r_e^\mu \hat r_e^\nu \P_e \bigg]+ \mathcal O((\bm q\ell_e)^3).\label{eq:single-band projected electron density operator}
\end{align}

Substituting the expanded density operators into~\cref{eq:CF single-particle Hamiltonian (full)} and absorbing the combined Gaussian factor into the Coulomb potential $V_{\bm q}\mapsto\widetilde{V}_{\bm q}\equiv V_{\bm q}e^{-\frac{1}{4}\bm q^2(\ell_v^2+\ell_e^2)}\equiv V_{\bm q}e^{-\lambda^2\bm q^2}$, we obtain the following matrix elements:
\begin{align*}
	H^P_0(\bm k,z_v;\bm k',z_v') &\simeq \frac{-c^2}{2}\sum_{\bm q}\widetilde{V}_{\bm q}\Big\langle\psi_{\bm k},z_v\Big| 2\left(1+\frac{1}{4}(\bm q\ell_e)^2\right)\cdot \P_e\otimes \P_{v,\text{LLL}} \nonumber \\
	& \qquad - q_\alpha q_\beta\bigg[ \P_e \otimes \hat R_{v,\alpha}\hat R_{v,\beta} + \P_e\hat r_{e,\alpha}\hat r_{e,\beta}\P_e \otimes \P_v - 2 \P_e\hat R_{e,\alpha} \P_e \otimes \hat R_{v,\beta}\bigg]\Big|\psi_{\bm k'},z_v'\Big\rangle.
\end{align*}
Now if we further assume $V_{\bm q}$, or $\widetilde{V}_{\bm q}$ to be \emph{isotropic}, then only components $\sim\frac{1}{2}\bm q^2\delta_{\mu\nu}$ survive after the angular integration, giving rise to
\begin{align}
	H_0^P(\bm k,z_v) &= -c^2\sum_{\bm q} \widetilde{V}_{\bm q}\bigg\{\left(1+\frac{1}{4}(\bm q\ell_e)^2\right) - \frac{1}{4}\bm q^2 \bigg[\langle\psi_{\bm k}|\P_e\hat{\bm r}_e^2 \P_e|\psi_{\bm k}\rangle + \langle z_v|\hat{\bm R}_v^2|z_v\rangle - 2 \langle\psi_{\bm k}|\P_e \hat{\bm r}_e \P_e|\psi_{\bm k}\rangle\cdot \langle z_v|\hat{\bm R}_v|z_v\rangle\bigg]\bigg\}\nonumber\\
	&=\text{const.}+\frac{c^2}{4}\sum_{\bm q} \widetilde{V}_{\bm q}\bm q^2 \bigg[\langle\psi_{\bm k}|\P_e\hat{\bm r}_e^2 \P_e|\psi_{\bm k}\rangle + \langle z_v|\hat{\bm R}_v^2|z_v\rangle - 2 \langle\psi_{\bm k}|\P_e \hat{\bm r}_e \P_e|\psi_{\bm k}\rangle\cdot \langle z_v|\hat{\bm R}_v|z_v\rangle\bigg].
\end{align}
This expression can be further simplified by noting that projecting the electron position operators to a Bloch band gives~\cite{roy2014band}
\begin{equation}\label{eq:band projection of single electron position operator}
	\P_e\hat r_e^\mu \P_e = \sum_{\bm k}|\psi_{\bm k}\rangle\mathcal D_\mu\langle\psi_{\bm k}|
\end{equation}
and
\begin{equation}\label{eq:band projection of electron position operator dot}
	\P_e \bm r_e^2 \P_e =\sum_{\bm k}|\psi_{\bm k}\rangle\Big[4\mathcal D \bar{\mathcal D} + (\Omega+\tr\mathcal G)\Big]\langle\psi_{\bm k}|
\end{equation}
with the covariant derivative $\mathcal{D}_\mu = i\partial_{k_\mu} + A_\mu(\bm k)$ and its complex $\mathcal{D} = (\mathcal{D}_x - i\mathcal{D}_y)/2$.
For the vortex guiding-center operators projected to its coherent states, it is helpful to recall that by definition $\hat R_v|z_v\rangle = z_v|z_v\rangle$ and $\hat{\bar R}_v|z_v\rangle = \hat{\bar z}_v|z_v\rangle$.
Here the operator $\hat{\bar z}_v$ can be set to $\hat{\bar z}_v=-2\ell_v^2\partial_{z_v}$ to obey the vortex guiding-center algebra $[\hat R_{v,x},\hat R_{v,y}]=i\ell_v^2$. So
we have, up to some constants,
\begin{equation*}
	P^v\hat{\bm R}_v^2 P^v = P^v\hat R_v \hat{\bar R}_v P^v = \sum_{z_v}|z_v\rangle z_v\hat{\bar z}_v\langle z_v|,
\end{equation*}
and
\begin{equation*}
	\bm{\mathcal D}\cdot\langle\hat{\bm R}_v\rangle\equiv \mathcal D\langle\hat R\rangle + \bar{\mathcal D}\langle\hat{\bar R}_v\rangle=\mathcal D z_v + \bar{\mathcal D}\hat{\bar z}_v.
\end{equation*}
As a result, $H^P_0(\bm k,z_v)$ can be summarized into a neat form up to some constant, reading
\begin{align}
	H_0^P(\bm k,z_v) &= \left(\frac{c^2}{4}\sum_{\bm q} \widetilde{V}_{\bm q}\bm q^2\right) \bigg[\langle\psi_{\bm k}|\P_e\hat{\bm r}_e^2 \P_e|\psi_{\bm k}\rangle + \langle z_v|\hat{\bm R}_v^2|z_v\rangle - 2 \langle\psi_{\bm k}|\P_e \hat{\bm r}_e \P_e|\psi_{\bm k}\rangle\cdot \langle z_v|\hat{\bm R}_v|z_v\rangle\bigg]\nonumber\\
	&=\frac{\hbar^2}{2m_*\ell_e^2\ell_v^2}\bigg[\Big(4\mathcal D\bar{\mathcal D} + (\tr\mathcal G+\Omega)\Big) + z_v\hat{\bar z}_v -2\Big(\mathcal D z_v + \bar{\mathcal D}\hat{\bar z}_v\Big)\bigg] \nonumber\\
	&=\frac{\hbar^2}{2m_*\ell_e^2\ell_v^2}\bigg[(\tr\mathcal G+\Omega) + \Big(2\mathcal D - z_v\Big)\Big(2\bar{\mathcal D} - \hat{\bar z}_v\Big)\bigg],
\end{align}
where the Coulomb-interaction-driven CF effective mass is defined as
\begin{equation}\label{eq:CF mass (supp)}
	\frac{\hbar^2}{2m_*\ell_e^2\ell_v^2} \equiv \frac{c^2}{4}\sum_{\bm q}\widetilde{V}_{\bm q}\bm q^2,\quad\text{or}\quad \frac{\hbar^2}{2m_*} \equiv \frac{\ell_e^4}{4}\sum_{\bm q}\widetilde{V}_{\bm q}\bm q^2.
\end{equation}
The magnetic lengths $\ell_e$ and $\ell_v$ are introduced to match the LLL result.

While the small-$\bm q$ expansion provides transparent analytic structure, all our numerical results are computed using the nonperturbative \emph{GMP expansion} over $N_s$ number of $\bm G_{\text{std}}$ vectors (see the next section for details):
\begin{equation}
	\bar\rho_{e,\bm q} = \sum_{\bm G_{\text{std}}} c(\bm q, \bm G_{\text{std}}) \rho^\text{GMP}_{e,[\bm q] + \bm G_{\text{std}}}
\end{equation}
for the full-$q$ Hamiltonian Eq.~\eqref{eq:CF single-particle Hamiltonian (full)}, without resorting to a small-$\bm q$ expansion or assuming any specific form for a regulator in $\widetilde{V}_{\bm q}=V_{\bm q}e^{-\lambda^2\bm q^2}$.
Effectively, this approach decomposes the FCI problem into a collection of FQHE-like problems, each evaluated under a vortex-LLL-form-factor-screened potential $\widetilde{V}_{\bm q}=V_{\bm q}e^{-(\bm q\ell_v)^2/4}$, weighted by the electronic GMP expansion coefficients $c(\bm q,\bm G_e)$. As such, our method rigorously accounts for all lattice-specific effects and band-projected electron form factors that are crucial for determining FCI stability.

\section{GMP Expansion}
For a $N_{\alpha,1}\times N_{\alpha,2}$ finite-size sample of species $\alpha=e,v,R,\eta$ under $N_{s,\alpha}\equiv N_{\alpha,1}\times N_{\alpha,2}$ number of flux quantum over a torus,
we can always prepared $N_{s,\alpha}$ number of independent magnetic Bloch states,  denoted as $\{|\psi^{\text{LLL}}_{\bm k}\rangle\}$, to respect all lattice symmetry, including magnetic translation symmetry and magnetic rotation symmetry following Jian and Qi's strategy in Ref.~\cite{jian2013crystal}, either from Haldane and Rezayi's periodic Laughlin wavefunction~\cite{haldane1985periodic} as we chose in Ref.~\cite{hu2024hyperdeterminants}, or Haldane's modular-invariant modified Weierstrass sigma function~\cite{haldane2018modular,haldane2018origin}. Here $\alpha=e,v,R,\eta$ represents the guiding-center coordinates of electrons, guiding-center coordinates of vortices, and guiding-center and cyclotron coordinates of CFs.

For convenience, we will ignore the species index $\alpha=e,v,R,\eta$ below, although one should to be cautious about the species-dependent quantities, including magnetic lengths $\ell_\alpha$, lattice constants $\bm a_{\alpha,\mu}$, sample sizes $N_{\alpha,\mu}$, and the signs of charges. The details of these quantities can be found in our previous paper Ref.~\cite{hu2024hyperdeterminants}.



\subsection{$N_s^2$ Number of LLL Density Operators}
The scattering momentum for the LLL density operator over the finite-size lattice generally takes the form of $\bm q\equiv l_1\frac{\bm G_1}{N_1} + l_2\frac{\bm G_2}{N_2}$ for $l_i\in\mathbb{Z}$. Let us introduce the $N_s^2$ number of the \emph{standard scattering momentum} $\bm q_{\text{std}}$, by constraining each $l_i\in\mathbb Z$ within $[0,N_s)$ that
\begin{equation*}
	\bm q_{\text{std}}\equiv l_1\frac{\bm G_1}{N_1} + l_2\frac{\bm G_2}{N_2},\quad l_i\in\{0,1,\ldots,N_s-1\}.
\end{equation*}

Any scattering momentum $\bm q$ can be uniquely decomposed as $\bm q\equiv[\bm q]+\{\bm q\}$, with $[\bm q]\in\text{BZ}$ the component folded back to the Brillouin zone (BZ), and $\{\bm q\}$ the needed shifts of multiples of BZ. So for this $\bm q_{\text{std}}$, we have the decomposition
\begin{align*}
	\bm q_{\text{std}}&\equiv[\bm q] +\bm G_{\text{std}}\\
	&\equiv[\bm q] +l_{\text{std},1}\bm G_1+l_{\text{std},2}\bm G_2,\quad l_{\text{std},1}\in\{0,1,\ldots,N_2-1\}\quad\text{and}\quad l_{\text{std},2}\in\{0,1,\ldots,N_1-1\}.
\end{align*}
All allowed $\{\bm G_{\text{std}}\}$ constitute a parallelogram $\mathcal Q_{\text{std}}$ named as the \emph{standard region} enclosed by two sides: $N_2\bm G_1$ and $N_1\bm G_2$, which is much larger than the original BZ. Clearly there are $N_2\times N_1=N_s$ number of allowed $\bm G_{\text{std}}$, equal to the number of points within the BZ. So there are totally $N_s^2$ number of standard scattering momentum $\bm q_{\text{std}}$.

In the same spirit, we can uniquely decompose an arbitrarily large scattering momentum $\bm q$ as some standard scattering momentum $\bm q_{\text{std}}$, together with with a suitable shifts of multiples of the standard region $\mathcal Q_{\text{std}}$, following
\begin{equation}\label{eq:decomposition of scattering momentum to standard region}
	\bm q\equiv \bm q_{\text{std}} +\bm G_{\text{shift}} \equiv \bm q_{\text{std}}+ m_1(N_2\bm G_1) + m_2(N_1\bm G_2),\quad m_i\in\mathbb{Z}.
\end{equation}

The sum form of the GMP algebra (magnetic translation algebra) satisfied by the LLL density operators $\rho_{\bm q}^{\text{LLL}}\equiv e^{-i\bm q\cdot\hat{\bm R}}$
then ensures that, there are only $N_s^2$ number of independent LLL density operators for the scattering momentum living within the standard region $\bm q\in\mathcal Q_{\text{std}}$. Precisely,
\begin{itemize}
	\item Any LLL density operator carrying momentum \emph{outside} the standard region is uniquely determined by the $N_s^2$ number of the LLL density operator \emph{inside} the standard region, with a phase factor analytically known.
\end{itemize}
In practice, we will use this fact extensively when constructing the preferred-charge density operator for the full-$q$ calculation of the CF Hamiltonian.

\subsection{GMP Expansion for Fermionic Bilinears}
A fermionic bilinear carrying momentum-$\bm q$ defined over the finite-size sample, such as the band-projected density operator we are focusing in the main text, expressed under bloch basis, takes the form (we will omit the band index)
\begin{equation*}
	A_{\bm q}\equiv\sum_{\bm k\in\text{BZ}}\mathscr A_{\bm q}([\bm k]) c_{[\bm k+\bm q]}^\dagger c_{\bm k}\equiv\sum_{\bm k\in\text{BZ}}\langle\psi_{[\bm k+\bm q]}|\mathscr A_{\bm q}|\psi_{\bm k}\rangle c_{[\bm k+\bm q]}^\dagger c_{\bm k}.
\end{equation*}
With $\bm q$ fixed, the resulting momentum folded back to the BZ $[\bm k+\bm q]$ is immediately known once the initial momentum $\bm k\in\text{BZ}$ is given. So the number of independent fermionic bilinears carrying fixed momentum-$\bm q$ equals to the number of Bloch states within BZ, denoted as $N_{\#[\bm k]}$.

Identify the $N_{\#[\bm k]}$ sites of the momentum grid as the zeros of some LLL wavefunction $\psi_{\text{LLL},\bm k}(\bm r)$ defined on a torus threaded by $N_s=N_{\#[\bm k]}$ flux quanta, we can construct another fermionic bilinear using the same bloch-band creation/annihilation operators, but now mixed with the LLL density operator matrix element carrying the same momentum-$\bm q$:
\begin{equation*}
	\rho_{\bm q}^{\text{GMP}}\equiv \sum_{\bm k\in\text{BZ}} \varrho_{\bm q}^{\text{GMP}}([\bm k]) c_{[\bm k+\bm q]}^\dagger c_{\bm k}\equiv\sum_{\bm k\in\text{BZ}}\langle\psi_{\text{LLL},[\bm k+\bm q]}|e^{i\bm q\cdot\hat{\bm R}}|\psi_{\text{LLL},\bm k}\rangle c_{[\bm k+\bm q]}^\dagger c_{\bm k}.
\end{equation*}
We add a superscript ``GMP'' \cite{girvin1986magneto} to the constructed fermionic bilinear to emphasize that it also satisfies the GMP algebra, so still there are at most $N_s^2$ number of independent GMP operators. The reason for introducing this new fermionic bilinear is that, for a fixed scattering momentum-$\bm q$, i.e. fixed BZ-folded scattering momentum $[\bm q]$, there are also $N_s$ number of independent $\bm G_{\text{std}}$ within the standard region $\mathcal Q_{\text{std}}$, and thus $N_s$ number of independent GMP operators $\varrho^{\text{GMP}}_{\bm q_{\text{std}}}$ spanned by $\bm q_{\text{std}}\equiv[\bm q]+\bm G_{\text{std}}$.

Now we are ready to write down the operator-level expansion, known as the \emph{GMP expansion}~\cite{murthy2012hamiltonian,hu2024hyperdeterminants}:
\begin{equation}\label{eq:GMP expansion with fixed momentum}
	A_{\bm q}\equiv\sum_{\bm G_{\text{std}}} c(\bm q,\bm G_{\text{std}})\rho^{\text{GMP}}_{[\bm q]+\bm G_{\text{std}}},
\end{equation}
with the expansion coefficient $c(\bm q,\bm G_{\text{std}})$ waited to be determined. \Cref{eq:GMP expansion with fixed momentum} can be established, as long as the following relation is true for every $\bm k\in\text{BZ}$:
\begin{equation*}
	\sum_{\bm G_{\text{std}}} c(\bm q,\bm G_{\text{std}})\varrho^{\text{GMP}}_{[\bm q]+\bm G_{\text{std}}}([\bm k])\equiv\mathscr A_{\bm q}([\bm k]).
\end{equation*}
Here, the LHS is matrix elements evaluated under the constructed magnetic Bloch basis, whereas the RHS is the matrix element evaluated under the Bloch basis of the system.

A useful property of the constructed GMP operator carrying the general scattering momentum
\begin{equation*}
	\bm q\equiv\bm q_{\text{std}}+\bm G_{\text{shift}}\equiv [\bm q]+\bm G_{\text{std}}+\bm G_{\text{shift}},
\end{equation*}
is that, it is \emph{traceless} for any scattering momentum except when $\bm q_{\text{std}}$ is vanishing, or when $\bm q=m_1(N_2\bm G_1)+m_2(N_1\bm G_2)$ for some integer multiples $m_i\in\mathbb Z$:
\begin{equation}\label{eq:trace of GMP operator}
	\Tr\Big[\varrho^{\text{GMP}}_{\bm q}\Big]\equiv\sum_{\bm k}\langle\psi_{\text{LLL},\bm k}|e^{-i\bm q\cdot{\bm R}}|\psi_{\text{LLL},\bm k}\rangle=N_s\delta_{\bm q_{\text{std}},\bm 0},
\end{equation}
and \emph{orthogonal} to each other in the sense of trace, under the help of the GMP algebra:
\begin{equation}\label{eq:trace orthogonality of two GMP operators}
	\Tr\Big[\varrho^{\text{GMP}}_{\bm q_1}\varrho^{\text{GMP}}_{\bm q_2}\Big]=\Tr\Big[e^{i\frac{\ell^2}{2}\bm q_1\wedge\bm q_2}\varrho^{\text{GMP}}_{\bm q_1+\bm q_2}\Big]=N_s\delta_{(\bm q_1+\bm q_2)_{\text{std}},\bm 0}.
\end{equation}
So after multiplying the other GMP matrix element $\varrho^{\text{GMP}}_{-[\bm q]-\bm G'_{\text{std}}}([\bm k+[\bm q]+\bm G'_{\text{std}}])=\varrho^{\text{GMP}}_{-[\bm q]-\bm G'_{\text{std}}}([\bm k+\bm q])$ on both sides for some $\bm G'_{\text{std}}$ and taking the trace, one immediately end up with the coefficient:
\begin{equation}
	c(\bm q,\bm G_{\text{std}})\equiv\frac{1}{N_s}\Tr\Big[\mathscr A_{\bm q}\varrho^{\text{GMP}}_{-[\bm q]-\bm G_{\text{std}}}\Big].
\end{equation}
Note: the neat notation $\Tr[\cdots]$ here only ensures the momentum-conservation. For the operator to be traced over, it can be evaluated under either the magnetic bloch states or the original bloch states, depending on the domain of itself. Applying this GMP expansion to the band-projected density operators, Eq.~(11) in the main text is then established.

This GMP expansion can be easily generalized to arbitrary fermionic bilinears $A\equiv\sum_{\alpha\beta} \mathscr A_{\alpha\beta} c_\alpha^\dagger c_\beta$ with $N_s^2$ number of independent matrix elements $\mathscr A_{\alpha\beta}$, where $\alpha,\beta=1,2,\ldots,N_s$ label the single-particle states within the finite-size sample. In this case, one only needs to make use of the full set of the $N_s^2$ number of LLL density operators $\varrho^{\text{GMP}}_{\bm q_{\text{std}}}$ to derive the operator-level expansion.

\section{Extended Data}
\subsection{tMoTe$_2$ Model at $\nu=-3/5$}
To see the generality of our theory to other fillings, we also compute the CF phase diagram and ED phase diagram for tMoTe$_2$ model at filling $\nu=-3/5$.

Using the same treatment in the main text, we take a particle-hole transformation for the single-band projected model to convert the $\nu=-3/5$ hole-filling states of tMoTe$_2$ to the equivalent $\nu=2/5$ electron-filling states, by squeezing our an extra single-particle term absorbed to the original electron dispersion. This operation helps to reduce the number of filled CF bands for our full-$q$ calculation. Following Jain's sequence $\nu=\frac{p}{2sp+1}$, this electronic filling has $p=2$, so \emph{two} CF bands get fully filled, and the ``CF gap'' is measuring the gap between the second (highest filled) and third (lowest empty) CF bands.
\begin{figure}[htb!]
	\centering
	\includegraphics[width=0.9\linewidth]{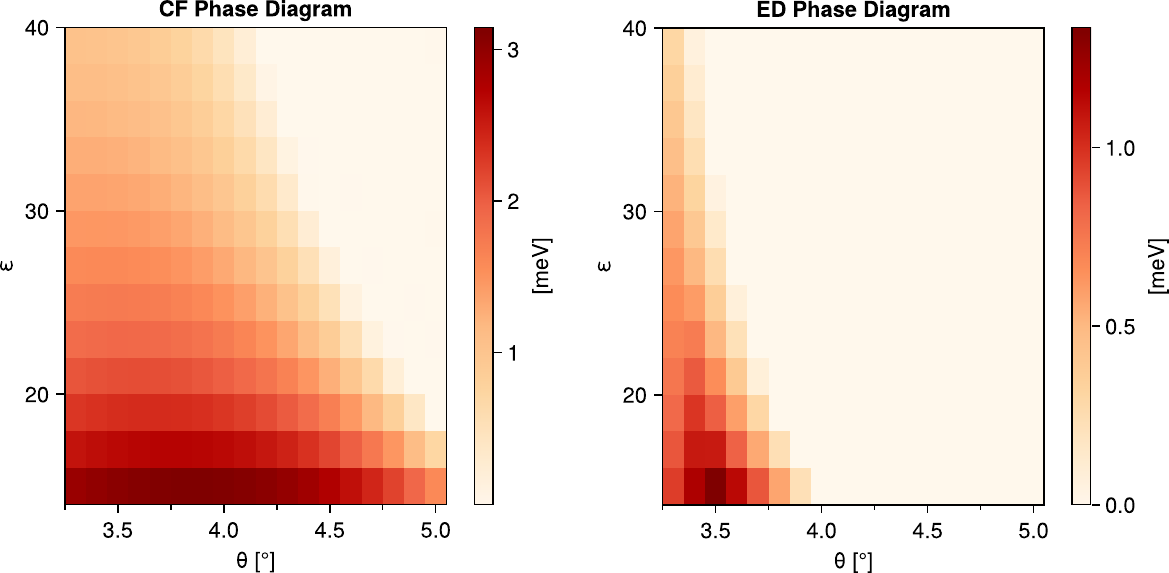}
	\caption{\textbf{CF Mean-field Phase Diagram and ED Phase Diagram for tMoTe$_2$ Model at $\nu=-3/5$}.}
	\label{fig:combined_tMoTe2_CF_phase_diagram}
\end{figure}

In~\cref{fig:combined_tMoTe2_CF_phase_diagram}, we calculated the full-$q$ CF spectra for the $5\times 5$ electronic sample and compare the CF phase diagram with the $5\times4$ ED phase diagram. Comparing with the CF phase diagram obtained at $\nu=-2/3$ in the main text, the CF gap size is smaller in the current case, as the $\nu=-3/5$ state is the daughter state of the $\nu=-2/3$ state. And the CF phase diagram still captures the overall trend of the FCI many-body gap evolution, and the location of the maximal gap.

\subsection{Modified Landau Level Model at $\nu=1/3$}
To further demonstrate the generality of our full-$q$ CF theory, we also computed the CF phase diagram and ED phase diagram for a totally different toy model by mixing the lowest two Landau levels (LLs) $\{|n=0\rangle,|n=1\rangle\}$ under a periodic potential $V_{pp}(\hat{\bm r}_e)$ respecting $C_4$ rotation symmetry, known as the modified LLL model~\cite{murthy2012hamiltonian,hu2024hyperdeterminants}.

Basically, the single-particle Hamiltonian is obtained by projecting such $C_4$ symmetric periodic potential $V_{pp}(\hat{\bm r}_e)\equiv\sum_{\bm G_e}V_{pp}(\bm G_e) e^{i\bm G_e\cdot\hat{\bm r}_e}$ onto the Landau level basis. Truncating the potential to the first-shell components only, with $V_{pp}(\bm G_e)=V_{10}/2$ for $\bm G_e=\pm\bm G_{e,i}$ and $i=1,2$, and manually inserting a LL spacing $\omega$ between the two LLs, we have \cite{hu2024hyperdeterminants}
\begin{align}\label{eq:MLL single-particle Hamiltonian}
	H &= \sum_{n,m\in\{0,1\}}\sum_{\bm G_e\in\text{first shell}}|n\rangle\langle n|V_{pp}(\bm G_e) e^{i\bm G_e\cdot\hat{\bm r}_e}|m\rangle\langle m| + \omega |n=1\rangle\langle n=1| \nonumber\\
	&=
	\begin{bmatrix}
		\frac{\widetilde{V}}{\sqrt{\pi}}(\cos k_{e,x}+\cos k_{e,y}) & \widetilde{V}(-i\sin k_{e,x}-\sin k_{e,y})\\[0.7em]
		\widetilde{V}(i\sin k_{e,x}-\sin k_{e,y}) & \omega+\frac{\pi-1}{\sqrt{\pi}}\widetilde{V}(\cos k_{e,x}+\cos k_{e,y})
	\end{bmatrix},\quad\text{with}\quad\widetilde{V}\equiv e^{\pi/2}\sqrt\pi V_{10}.
\end{align}
The MLL model~\cref{eq:MLL single-particle Hamiltonian} appears as a massive Dirac model that stays in a topological regime with Chern number $C=-1$ (note the LL orbital itself also carries Chern number $-1$), as long as the periodic potential satisfies $V_{10}\leq V_{10,c}\equiv\frac{e^{\pi/2}}{2\pi}\omega\approx 0.766\omega$.

We consider the UV softened Coulomb interaction $V_{\bm q_e}\equiv\frac{2\pi e^2}{\epsilon|\bm q_e|} e^{-\beta^2|\bm q_e\ell_e|^2}$ with the dimensionless parameter $\beta=1.0$. Here effective magnetic length $\ell_e^2 \equiv \hbar/(eB_{\text{eff}})$ is set by the Berry flux density $B_{\text{eff}}\equiv h/(e A_{\text{u.c.}})$. Note: this potential only regularize the UV behavior $q\rightarrow\infty$ but stays to be long-ranged in the IR limit. We then project this potential onto the bands solved from~\cref{eq:MLL single-particle Hamiltonian} and fill the lowest band at the fraction $\nu=1/3$.
\begin{figure}[!htb]
	\centering
	\includegraphics[width=0.9\textwidth]{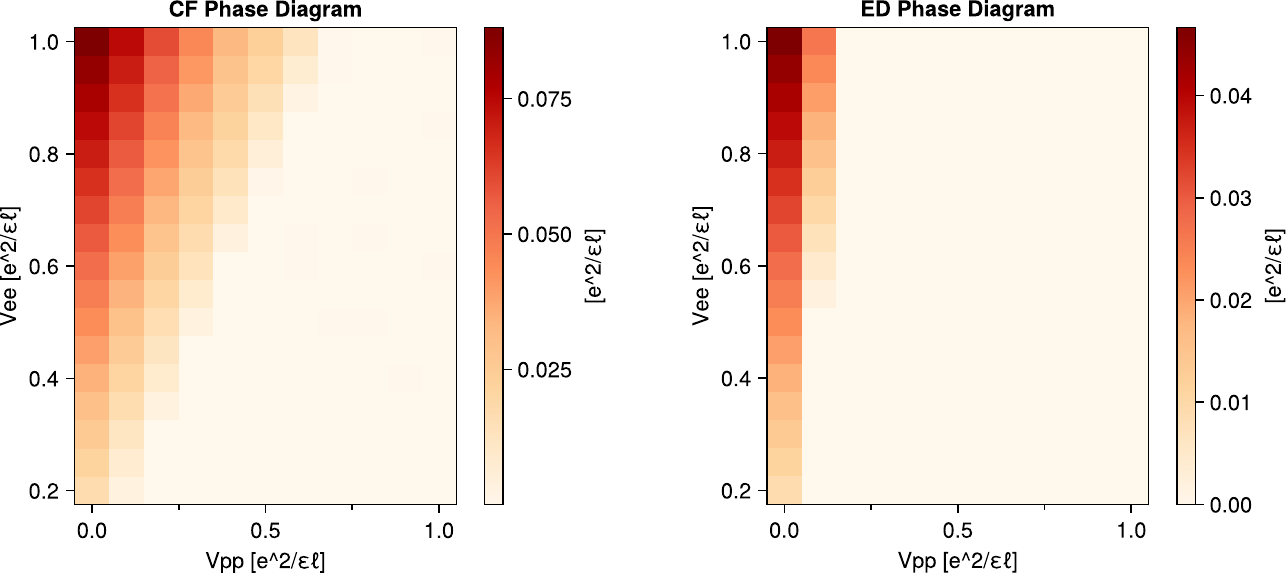}
	\caption{\textbf{CF Mean-field Phase Diagram and ED Phase Diagram for Modified Landau Level Model at $\nu=1/3$}.}
	\label{fig:combined_MLL_CF_phase_diagram}
\end{figure}

In~\cref{fig:combined_MLL_CF_phase_diagram}, we fix the LL spacing $\omega=5.0$ unit of $e^2/(\epsilon\ell_e)$, and tune the periodic potential strength $V_{10}$ from $0.0$ to $1.0$ unit of $e^2/(\epsilon\ell_e)$ (which is far away from the critical value $V_{10,c}\approx0.766\omega=3.83$ unit of $e^2/(\epsilon\ell_e)$ to trigger the  single-particle topological phase transition). We calculated the full-$q$ CF spectra for a $6\times 6$ electronic sample and compare the CF phase diagram with the $6\times 4$ ED phase diagram. Remarkably, the CF phase diagram still captures the overall trend of the FCI many-body gap evolution, and the location of the maximal gap.

\bibliography{ref}